\documentclass[conference]{IEEEtran}
\IEEEoverridecommandlockouts
\usepackage{multirow}
\usepackage{boldline,multirow}
\usepackage{cite}
\usepackage{amsmath,amssymb,amsfonts}
\usepackage{algorithm}
\usepackage{algpseudocode}
\usepackage{graphicx}
\usepackage{textcomp}
\usepackage{xcolor}
\usepackage{booktabs}
\usepackage{seqsplit}
\usepackage{tabularx}
\usepackage{ragged2e}

\usepackage{seqsplit}
\newcolumntype{Y}{>{\RaggedRight\arraybackslash}X}
\newenvironment{scenario}{\begin{list}{}{\leftmargin=1.5em}\item[]}{\end{list}}
\usepackage{atbegshi}
\usepackage{amsmath}
\usepackage{array} 
\usepackage{subcaption}
\usepackage{bm}
\usepackage{breakurl}
\usepackage{mathrsfs}   
\usepackage{calligra}   
\usepackage{url}
\usepackage{hyperref}
\usepackage{threeparttable}
\usepackage{makecell}
\usepackage{algpseudocode}
\algrenewcommand\algorithmiccomment[1]{\hfill$\triangleright$~#1} 
\usepackage[normalem]{ulem}
\useunder{\uline}{\ul}{}

\newcolumntype{C}[1]{>{\centering\arraybackslash}p{#1}}
\def\BibTeX{{\rm B\kern-.05em{\sc i\kern-.025em b}\kern-.08em
    T\kern-.1667em\lower.7ex\hbox{E}\kern-.125emX}}

\newcommand{\field}[1]{\emph{#1}}   
\newcommand{\param}[1]{\emph{#1}}   
\hypersetup{breaklinks=true}
\usepackage{amsmath,amssymb,amsfonts}
\usepackage{amssymb} 
\usepackage{pifont}  
\newcommand{\xmark}{\ding{55}} 
\usepackage{pdfpages}
\usepackage{graphicx}
\usepackage{textcomp}
\usepackage{xcolor}
\usepackage{array,tabularx}
\newcolumntype{L}[1]{>{\raggedright\arraybackslash}p{#1}}
\def\BibTeX{{\rm B\kern-.05em{\sc i\kern-.025em b}\kern-.08em
    T\kern-.1667em\lower.7ex\hbox{E}\kern-.125emX}}

\begin{document}
\title{A Protocol-Aware P4 Pipeline for MQTT Security and Anomaly Mitigation in Edge IoT Systems}

\author{
    \IEEEauthorblockN{
  Bui Ngoc Thanh Binh\IEEEauthorrefmark{1},
  Pham Hoai Luan\IEEEauthorrefmark{1},
  Le Vu Trung Duong\IEEEauthorrefmark{1},
  Vu Tuan Hai\IEEEauthorrefmark{2}\IEEEauthorrefmark{3}, 
  Yasuhiko Nakashima\IEEEauthorrefmark{1}
}
\IEEEauthorblockA{\IEEEauthorrefmark{1}Nara Institute of Science and Technology (NAIST), 8916-5 Takayama Science Town, Ikoma, Nara, Japan}
\IEEEauthorblockA{\IEEEauthorrefmark{2}University of Information Technology, Ho Chi Minh City, Vietnam}
\IEEEauthorblockA{\IEEEauthorrefmark{3}Vietnam National University, Ho Chi Minh City, Vietnam}
    Email: bui.ngoc\_thanh\_binh.bp6@naist.ac.jp

}


\maketitle

\begin{abstract}
MQTT is the dominant lightweight publish--subscribe protocol for IoT deployments, yet edge security remains inadequate. Cloud-based intrusion detection systems add latency that is unsuitable for real-time control, while CPU-bound firewalls and generic SDN controllers lack MQTT awareness to enforce session validation, topic-based authorization, and behavioral anomaly detection. We propose a P4-based data-plane enforcement scheme for protocol-aware MQTT security and anomaly detection at the network edge. The design combines parser-safe MQTT header extraction with session-order validation, byte-level topic-prefix authorization with per-client rate limiting and soft-cap enforcement, and lightweight anomaly detection based on KeepAlive and Remaining Length screening with clone-to-CPU diagnostics. The scheme leverages stateful primitives in BMv2 (registers, meters, direct counters) to enable runtime policy adaptation with minimal per-packet latency. Experiments on a Mininet/BMv2 testbed demonstrate high policy enforcement accuracy (99.8\%, within 95\% CI), strong anomaly detection sensitivity (98\% true-positive rate), and high delivery (\(>\)99.9\% for 100--5~kpps; 99.8\% at 10~kpps; 99.6\% at 16~kpps) with sub-millisecond per-packet latency. These results show that protocol-aware MQTT filtering can be efficiently realized in the programmable data plane, providing a practical foundation for edge IoT security. Future work will validate the design on production P4 hardware and integrate machine learning--based threshold adaptation.

\end{abstract}


\section{Introduction}

The Internet of Things (IoT) ecosystem continues to expand rapidly. IDC forecasts that by 2025 more than 55.9~billion connected devices will generate approximately 79.4~zettabytes of data \cite{idc2018digitization}. At the core of this growth is the Message Queuing Telemetry Transport (MQTT) protocol, now the de facto standard for lightweight publish–subscribe communication in resource-constrained environments. However, the rapid adoption of MQTT has exposed security weaknesses that threaten system integrity, confidentiality, and availability.

Recent assessments underscore the scope of the problem: over 47{,}000 MQTT brokers remain publicly accessible without authentication, while 98\% of IoT traffic is unencrypted and 57\% of devices exhibit medium- or high-severity vulnerabilities \cite{asgar2025mqtt,unit42iot2020}. In 2024 the risk sharpened with critical MQTT issues such as CVE-2024-6786 (path traversal) \cite{cve-2024-6786}, CVE-2024-31409 (wildcard exposure) \cite{cve-2024-31409}, and CVE-2024-31041 (DoS via null dereference) \cite{cve-2024-31041}, highlighting the need for protocol-aware edge enforcement. These application-layer vulnerabilities cannot be mitigated by traditional network defenses. Cloud-based intrusion detection adds latency incompatible with real-time IoT control (hundreds of milliseconds). CPU-bound firewalls cannot sustain deep packet inspection at line rate. L3/L4 SDN controls lack MQTT semantics, so they cannot enforce session validation, topic authorization, or behavioral screening. This gap motivates a programmable data-plane approach that couples protocol awareness with hardware-accelerated processing at the edge.

P4 enables custom parsing and stateful processing in network hardware while preserving wire-speed operation; registers, counters, meters, and clone-to-CPU provide fine-grained control without sacrificing throughput. Prior work has not fully leveraged P4 for MQTT security, often targeting generic firewalls or MQTT-SN and omitting session-order checks, byte-level topic authorization, and integrated anomaly screening.

We introduce a P4-based MQTT security pipeline that delivers protocol-aware enforcement and anomaly screening at the edge. The design includes a parser-safe path for variable-length fields (IPv4/TCP options, Remaining Length, topic strings) that drops malformed or evasive fragments; stateful ingress enforcement with explicit session-order validation, byte-level ternary topic-prefix ACLs with direct counters, and per-client soft limits with three-color metering; and lightweight anomaly screening via KeepAlive-gap and Remaining-Length heuristics, with suspicious traffic cloned to the control plane using preserved metadata.

On the BMv2/v1 model, the pipeline maintains a throughput exceeding 99.8\% delivery, with sub-millisecond latency per packet and achieves high enforcement precision.  The results indicate that awareness of the MQTT protocol can be effectively implemented in the programmable data plane; validation on production P4 hardware targets is reserved for future research.

\section{Related Work}
\subsection{MQTT Security and Intrusion Detection System at the Edge}
A significant amount of MQTT security research has concentrated on anomaly detection and attack mitigation through machine learning methodologies, generally implemented on backend servers or edge cloud clusters. Recent studies present one-class models and sophisticated traffic feature engineering, attaining elevated detection rates on public MQTT datasets. These approaches excel at identifying deviations from normal behavior with accuracy exceeding 95\%, yet they impose significant computational overhead and incur prohibitive latency measured in hundreds of milliseconds to seconds. This latency overhead renders ML-based systems unsuitable for real-time protection at resource-limited IoT edge gateways, where security decisions must occur at microsecond timescales. Innovative methodologies attempt to amalgamate real-time machine learning inference with edge infrastructure, however, they primarily depend on software processes and inadequately enforce measures within the network data plane, creating both a single point of failure and a scalability bottleneck as IoT deployments expand \cite{Asulba2025}.
\subsection{Programmable Data Plane P4 for Network Security}
Programmable data planes using P4 have emerged as a promising platform for implementing network security functions at line rate. P4 switches leverage lookup tables, direct counters, registers, and meters embedded in the pipeline to detect and mitigate attacks with low latency and high throughput. Recent comprehensive surveys systematize the capabilities, constraints, and design strategies of P4-enabled programmable switches, highlighting the critical role of stateful elements such as counters, registers, and meters in supporting real-time attack detection and rate limiting at high speeds. P4DDPI exemplified this potential by implementing deep packet inspection for DNS at line rate using stateful processing and recirculation techniques to achieve wire-speed throughput. However, these hardware capabilities introduce challenges related to limited memory and conditional branching, which must be carefully managed for efficient deployment \cite{alsabeh2022p4ddpi}.

\begin{table}[t]
\caption{Comparative Analysis of MQTT Security Approaches}
\label{tab:comparison}
\centering
\renewcommand{\arraystretch}{1.1}
\setlength{\tabcolsep}{2.5pt}

\makebox[\columnwidth][l]{%
\resizebox{\columnwidth}{!}{%
\begin{tabular}{lccccccc}
\hline
\textbf{Approach} & \textbf{Prot.} & \textbf{Sess.} & \textbf{Topic} & \textbf{Per-} & \textbf{Line-} & \textbf{Anom.} & \textbf{Deploy.} \\
 & \textbf{Aware} & \textbf{Order} & \textbf{ACL} & \textbf{Client} & \textbf{Rate} & \textbf{Det.} & \textbf{Loc.} \\
\hline
ML-based IDS & \checkmark & \xmark & \checkmark & \xmark & \xmark & \checkmark & Cloud \\
P4DDPI & \checkmark (DNS) & N/A & N/A & \checkmark & \checkmark & \checkmark & Edge \\
MQTT-SN P4 & \checkmark & \xmark & \xmark & \checkmark & \checkmark & \xmark & Edge \\
SDN-based Firewall & \checkmark & \xmark & \checkmark & \xmark & \xmark & \xmark & Ctrl. \\
\textbf{Our} & \textbf{\checkmark} & \textbf{\checkmark} & \textbf{\checkmark} & \textbf{\checkmark} & \textbf{\checkmark} & \textbf{\checkmark} & \textbf{Edge} \\
\hline
\end{tabular}%
}
}

\vspace{2pt}
\begin{minipage}{\columnwidth}
\scriptsize
Edge = Edge Switch, Ctrl. = Controller, Cloud = Backend/Cloud. “Sess. Order” = Session Order Enforcement, “Topic ACL” = Topic-Prefix Access Control, “Per-Client” = Per-Client Metering, “Line-Rate” = Line-Rate Processing, “Anom. Det.” = Anomaly Detection, “Deploy. Loc.” = Deployment Location.
\end{minipage}
\end{table}

\subsection{Programmable Data Plane for IoT/MQTT Security and Anomaly Monitoring}
Recent advancements in programmable data planes utilizing P4 have facilitated the offloading of packet processing tasks, traditionally managed by software, to network hardware, thus ensuring line-rate security enforcement and traffic management. A significant area of research emphasizes the integration of IoT protocol awareness, specifically MQTT and MQTT-SN, within the switch dataplane. Enhancements in MQTT-SN protocol processing utilizing P4 illustrate that certain broker functionalities, including topic matching and QoS 0 forwarding, can be transferred to the data plane, thereby diminishing latency and alleviating the burden on brokers. P4-based telemetry and direct counters have been used to fingerprint application-layer behavior and pipeline states for behavioral monitoring, enabling context-aware anomaly detection directly at the switch and enhancing traditional intrusion detection by reducing detection latency and response time \cite{banno2022p4}. Despite these advances, combining protocol-aware enforcement, behavioral telemetry, and lightweight anomaly filters for complex IoT protocols such as MQTT remains largely unresolved. This study addresses that gap by integrating parser-safe MQTT header extraction, client-specific stateful tracking, topic-prefix ACL enforcement, and heuristic anomaly detection within a P4-enabled edge switch pipeline.
\subsection{Comparison and Gaps}

MQTT security research has made progress, but few works enforce protocol-specifically for full MQTT traffic at the edge with true line-rate processing.  ML-based anomaly detection is accurate but slow due to backend processing, limiting real-time IoT response.  Although P4DDPI achieved line-rate deep packet inspection for DNS traffic, its architecture does not easily adapt to MQTT's stateful requirements: session lifecycle tracking (CONNECT/DISCONNECT), QoS enforcement, and hierarchical topic-based subscriptions.  MQTT-SN P4 implementations offload topic matching to the data plane to reduce broker load, but they only support binary MQTT-SN, lack session-order validation, and offer limited topic-based access control.  Traditional L3/L4 SDN firewalls cannot enforce application-layer policies like MQTT session state validation or topic authorization.

 We natively integrate protocol-aware enforcement, stateful session tracking, and heuristic anomaly detection in the P4 data plane without sacrificing line-rate throughput.  Parser-safe MQTT header extraction for variable-length fields and malformed packets, byte-level ternary topic-prefix ACLs for fine-grained hierarchical policies, explicit session-order enforcement for PUBLISH operations, and lightweight KeepAlive baseline tracking and per-client rate limiting are our contributions. In Table~\ref{tab:comparison}, we compare our approach to previous research.  We create a practical edge security architecture that aligns with emerging paradigms in edge-centric, semantically aware security by combining full MQTT protocol awareness, stateful session validation, and real-time behavioral tracking in a single P4 v1model/BMv2 pipeline.

\section {System Design}

\begin{figure}[t]
  \centering
  \hspace*{0\linewidth}%
  \makebox[\linewidth][l]{%
    \includegraphics[width=1\linewidth,page=1]{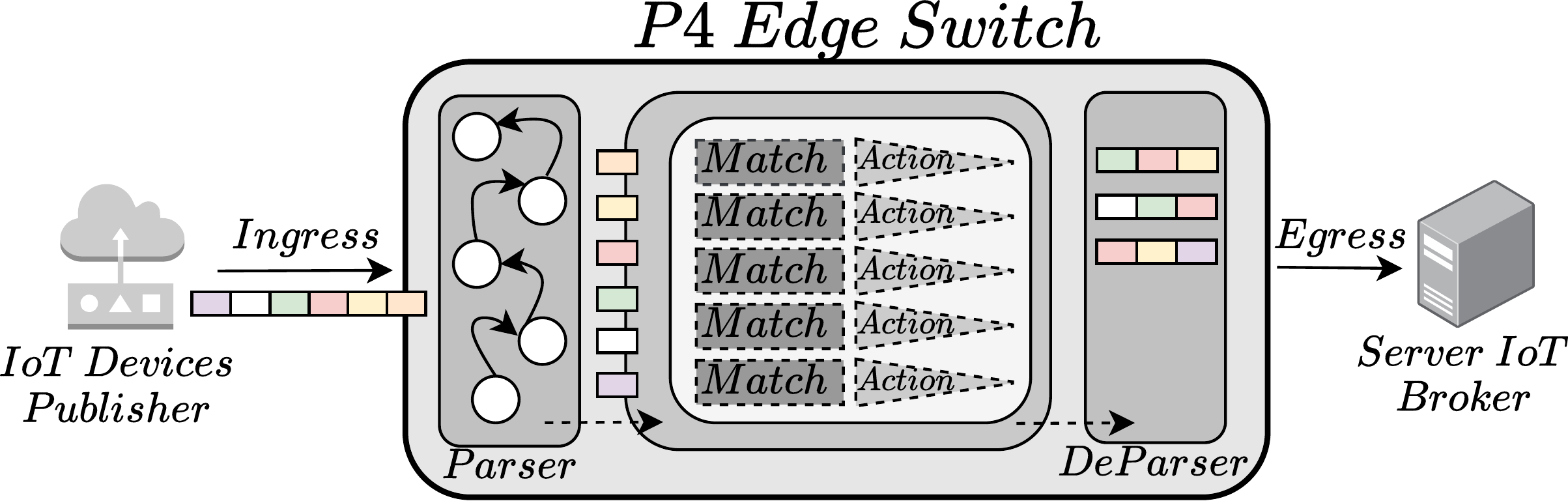}%
  }
  \caption{Edge MQTT security with protocol-aware P4 enforcement and Anomaly cloning to control-plane }
  \vspace{-0.4cm}
  \label{fig:arch}
\end{figure}

\subsection{Overall Architecture}

The P4 v1model pipeline runs on an edge switch between IoT publishers and the MQTT broker, enabling protocol-aware enforcement at line rate. It executes in a single forward pass across five stages: the parser extracts headers; ingress control enforces policies; the traffic manager performs cloning; egress applies any post-processing; and the deparser reconstructs the packet. Fig.~\ref{fig:arch} also shows the end-to-end packet flow. \textit{Parser.} The parser extracts Ethernet, IPv4, TCP, and MQTT. It skips IPv4/TCP options when \field{IHL} or \field{dataOffset} $>5$ and processes only first fragments (\field{fragOffset}$=0$). On destination port~1883 it branches by MQTT type: for Connect it captures the variable header including \field{KeepAlive}; for Publish it slices a 16-byte topic prefix for ACL matching. \textit{Ingress control.} Parsed metadata drives classification, a per-client index from the source address, counter updates, Connect before Publish validation, per-client soft-limit checks, lightweight anomaly heuristics (KeepAlive gap, Remaining Length), per-client metering, and ternary topic-prefix ACLs. Packets are tagged to Forward, Drop, or Clone, and metadata records the reason code, rule ID, and salient fields for diagnosis. \textit{Traffic manager / egress / deparser.} The traffic manager replicates packets marked for cloning to a CPU-facing port while originals continue downstream. Egress performs minimal post-processing as required, and the deparser emits valid headers in order. Forwarded and dropped packets stay on the fast path and do not interact with the control plane. \textit{Control plane (orthogonal).} Policies and parameters (such as \param{soft limit}, \field{KeepAlive} multiplier) are updated at runtime via table APIs without recompilation. The control plane reads direct counters and per-client registers and consumes cloned packets for telemetry, enabling real-time monitoring and threshold tuning without impacting data-plane performance.

\subsection{Protocol-Aware Parser and Header Extraction}

\begin{figure}[t]
  \centering
  \includegraphics[width=0.8\columnwidth]{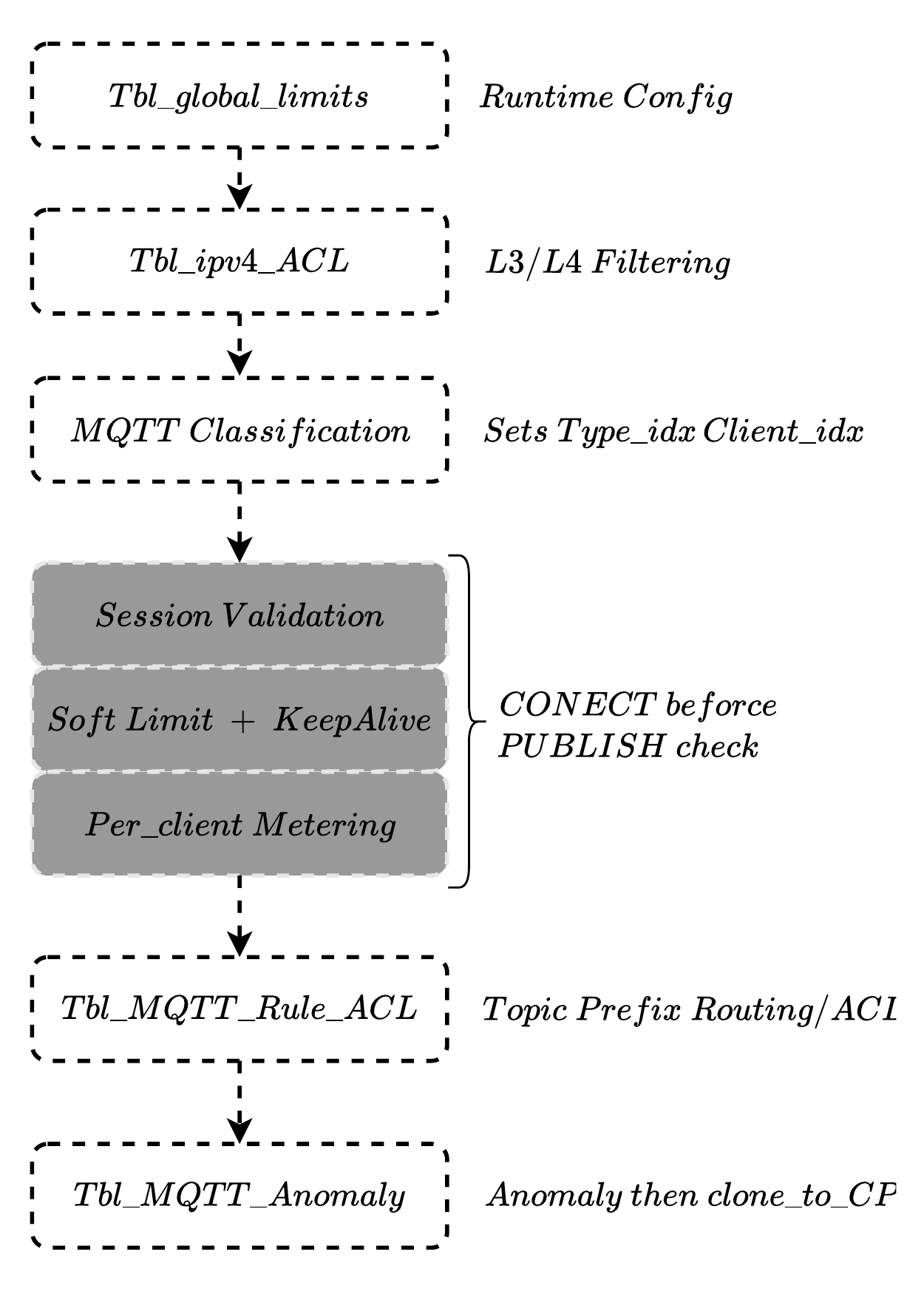}
  \caption{Simplified ingress-data flow illustrating table order and pipeline logic.}
  \label{fig:overview}
  \vspace{-0.5cm}
\end{figure}

Our parser enforces two safeguards against malformed packets: it intelligently skips options by computing IPv4/TCP option lengths and advancing the packet pointer to avoid parser exceptions, and it filters fragments by processing only first fragments (\(\mathit{fragOffset}=0\)) to prevent evasion.

\textit{Ethernet/IPv4/TCP parsing.}
The parser begins by extracting the Ethernet header to identify IPv4 packets. Upon detecting IPv4 (\(\text{etherType}=0x0800\)), it extracts the IPv4 header, including the Internet Header Length (\(\mathit{IHL}\)) field. If \(\mathit{IHL}>5\), the total option length is computed as

\begin{equation}
\mathrm{opt\_len}_{\mathrm{IPv4}} = (\mathit{IHL}-5)\times 4\,\mathrm{bytes}.
\label{eq:ipv4_optlen}
\end{equation}

Similarly, after extracting the TCP header, if the Data Offset (\(\mathit{doff}\)) exceeds \(5\), the parser advances by

\begin{equation}
\mathrm{opt\_len}_{\mathrm{TCP}} = (\mathit{doff}-5)\times 4\,\mathrm{bytes}.
\label{eq:tcp_optlen}
\end{equation}

This strategy avoids allocating large fixed-size option structures and prevents parser state machines from stalling on variable-length fields. Fragment filtering occurs at the IPv4 stage: the parser proceeds to TCP only if \(\mathit{fragOffset}=0\), preventing evasion via fragmented MQTT headers. Table~\ref{tab:variables} summarizes all variables, metadata fields, and register names referenced in our algorithm and throughout the pipeline description.

\begin{table}[t]
\caption{Variables and Field Definitions}
\label{tab:variables}
\centering
\footnotesize
\setlength{\tabcolsep}{3pt}
\renewcommand{\arraystretch}{1.05}
\begin{tabularx}{\columnwidth}{@{}lY@{}}
\toprule
\textbf{Variable} & \textbf{Definition / Usage} \\
\midrule
\texttt{rl\_b0}, \texttt{rl\_b1} & First two bytes of MQTT Remaining-Length for DoS detection \\
$t_0,\ldots,t_{15}$ & First 16~bytes of topic string for per-byte ternary ACL matching \\
\texttt{KeepAlive} & MQTT Connect timer interval (seconds) used for anomaly detection \\
\texttt{\seqsplit{reg\_session\_open[idx]}} & Session state flag (1~bit per client; 1 = connected) \\
\texttt{\seqsplit{reg\_keepalive\_s[idx]}} & Stored \texttt{KeepAlive} value (16~bit) \\
\texttt{\seqsplit{reg\_pkt\_total[idx]}} & Total packet counter (32~bit per client) \\
\texttt{\seqsplit{reg\_pkt\_per\_type[idx][type\_idx]}} & Per-type packet counter (4 $\times$ 32~bit per client) \\
\texttt{\seqsplit{reg\_last\_total[idx]}} & Baseline counter for KeepAlive gap \\
\texttt{\seqsplit{reg\_total[idx]}} & Current aggregated packet count \\
\texttt{idx} & Client index computed as $\mathit{srcAddr}\bmod 512$ \\
\texttt{pps\_factor} & Scaling constant for KeepAlive threshold (default 2) \\
\texttt{pub\_soft\_limit} & Max publish messages per client (default $20{,}000$) \\
\bottomrule
\end{tabularx}
\vspace{-0.3cm}
\end{table}

\textit{MQTT fixed header.} When the destination port is 1883, the parser extracts \(\mathit{type}\) (4~bits), \(\mathit{flags}\) (4~bits), and the first two bytes of Remaining Length (\texttt{rl\_b0}, \texttt{rl\_b1}). If \(\texttt{rl\_b1}=1\) (indicating a 3-byte Remaining Length and a potential DoS vector), the packet is flagged as suspicious.

\textit{Connect.} The parser reads the protocol-name length prefix, advances by that length, and then extracts the fixed fields: protocol level, flags, and \(\mathit{KeepAlive}\) (2~bytes), which are stored for anomaly detection.

\textit{Publish topic slicing.} The parser extracts the topic length, slices the first 16~bytes into per-byte fields (\(t_0\)–\(t_{15}\)) for ternary ACL matching, and skips any remaining bytes when the topic exceeds 16~bytes. This preserves the security-critical topic prefix while keeping memory usage bounded.

\subsection{Policy Enforcement Blocks}

The ingress pipeline enforces stateful policies at line rate in a single forward 
pass Fig.~\ref{fig:overview}, using runtime parameters fetched once from 
\texttt{tbl\_global\_limits} and a per-client index (Eq.~\ref{eq:client_idx}) 
to ensure consistent register and meter addressing. A coarse IPv4/TCP ACL 
(\texttt{tbl\_ipv4\_acl}) drops traffic that fails L3/L4 policy before any 
MQTT parsing.

Each packet's source address is hashed to a compact per-client index via modulo:

\vspace{-0.2cm}
\begin{equation}
\mathit{idx} = \mathit{srcAddr} \bmod 512,
\label{eq:client_idx}
\end{equation}

This $512$-entry address space balances collision probability against available 
per-client register footprint in BMv2. All per-client state (registers and meters 
indexed by $\mathit{idx}$) use this unified addressing scheme.

For permitted flows, MQTT messages are classified and bound to per-client state. A valid Connect with a negotiated \(\mathit{KeepAlive}\) opens the session and records the interval; any Publish observed without an open session is dropped (reason~180). Authorization is then enforced by a topic-prefix ACL (\texttt{tbl\_mqtt\_rule\_acl}) that requires Publish with QoS in \(\{0,1,2\}\), an authorized source subnet, and a match between the first 16~bytes of the topic and an approved prefix such as \texttt{device/sensor/*}. Rules maintain direct counters for visibility; non-matching traffic is dropped.

Work-conserving protections run alongside authorization. Per-type counters (tracking Publish, Subscribe, etc.) are aggregated per client; a soft cap is imposed on the Publish counter specifically. Exceeding \(\mathit{pub\_soft\_limit}\) (default \(20{,}000\) Publish messages per client) triggers drops (reason~181) to curb floods. A three-color meter provides hardware rate enforcement; packets marked RED are dropped (reason~150).

We track the most recent keep-alive event per client (CONNECT or PINGREQ) using a per-client timestamp register \texttt{reg\_last\_ka\_ts[idx]} initialized on the first CONNECT. On each ingress packet, we compute the elapsed time since the last keep-alive \(\Delta t\) as in Eq.~\ref{eq:ka_delta}:
\begin{equation}
\Delta t = \frac{\texttt{ingress\_ts} - \texttt{reg\_last\_ka\_ts}[\mathit{idx}]}{10^9}\; [\mathrm{s}],
\label{eq:ka_delta}
\end{equation}
where \texttt{ingress\_ts} is the per-packet ingress timestamp (ns) from standard metadata. A violation is raised when the condition in Eq.~\ref{eq:ka_violation} holds:
\begin{equation}
\Delta t > \gamma \cdot \texttt{KeepAlive},
\label{eq:ka_violation}
\end{equation}
where \(\gamma\) is a tolerance multiplier (default \(1.5\)) set via \texttt{tbl\_global\_limits}. Violations trigger reason~182 cloning to the control plane but do not reset the baseline, allowing offline analysis of sustained stalls. The check is skipped when \texttt{KeepAlive}{=}0 (per MQTT semantics), and only valid CONNECT or PINGREQ messages update \texttt{reg\_last\_ka\_ts[idx]}.

For Remaining-Length anomaly detection, let \(\mathrm{RL}\) denote the decoded Remaining Length; we flag and clone (reason~183) when the threshold is met $\mathrm{RL} \ge \theta_{\mathrm{RL}}$, where \(\theta_{\mathrm{RL}}\) is an adjustable threshold (default \(16{,}384\) bytes; for our experiments, \(131{,}072\) to prevent false positives with extensive sensor data). Deployments may optionally require that the RL encoding use more than three bytes to trigger flagging, further reducing false positives for legitimate but sizable packets.

To support diagnosis without disrupting the fast path, the pipeline preserves explanatory metadata with each decisive action. Specifically, it stamps the reason code, Table~\ref{tab:reasons}, the matched table and rule identifier (such as the \texttt{tbl\_mqtt\_rule\_acl} entry index), and salient parsing context (client index, MQTT type/QoS, first 16 topic bytes) into a dedicated metadata struct that travels with the packet. When cloning is invoked, \texttt{clone\_preserving\_field\_list} reproduces these fields verbatim for the control plane, enabling counter correlation and policy refinement while the original packet continues through egress at line rate.

\begin{table}[t]
\caption{Reason Codes for Packet Clone and Drop Actions}
\label{tab:reasons}
\centering
\small
\setlength{\tabcolsep}{3pt}
\renewcommand{\arraystretch}{1.05}
\begin{tabular}{@{}c p{.52\columnwidth} p{.34\columnwidth}@{}}
\toprule
\textbf{Code} & \textbf{Trigger condition} & \textbf{Pipeline stage} \\
\midrule
150 & Rate limit exceeded (RED meter) & Per-client metering \\
180 & Publish without session & Session validation \\
181 & Publish soft limit surpassed & Per-client soft limits \\
182 & \(\mathit{KeepAlive}\) violation  & Anomaly detection \\
183 & Remaining Length \(\geq\!3\)~bytes & Anomaly detection \\
\bottomrule
\end{tabular}
\vspace{-0.3cm}
\end{table}

\section {Implementation and Evaluation}
\subsection{Testbed Setup and Configuration}

\begin{figure}[t]
  \centering
  \includegraphics[height=1.2\linewidth]{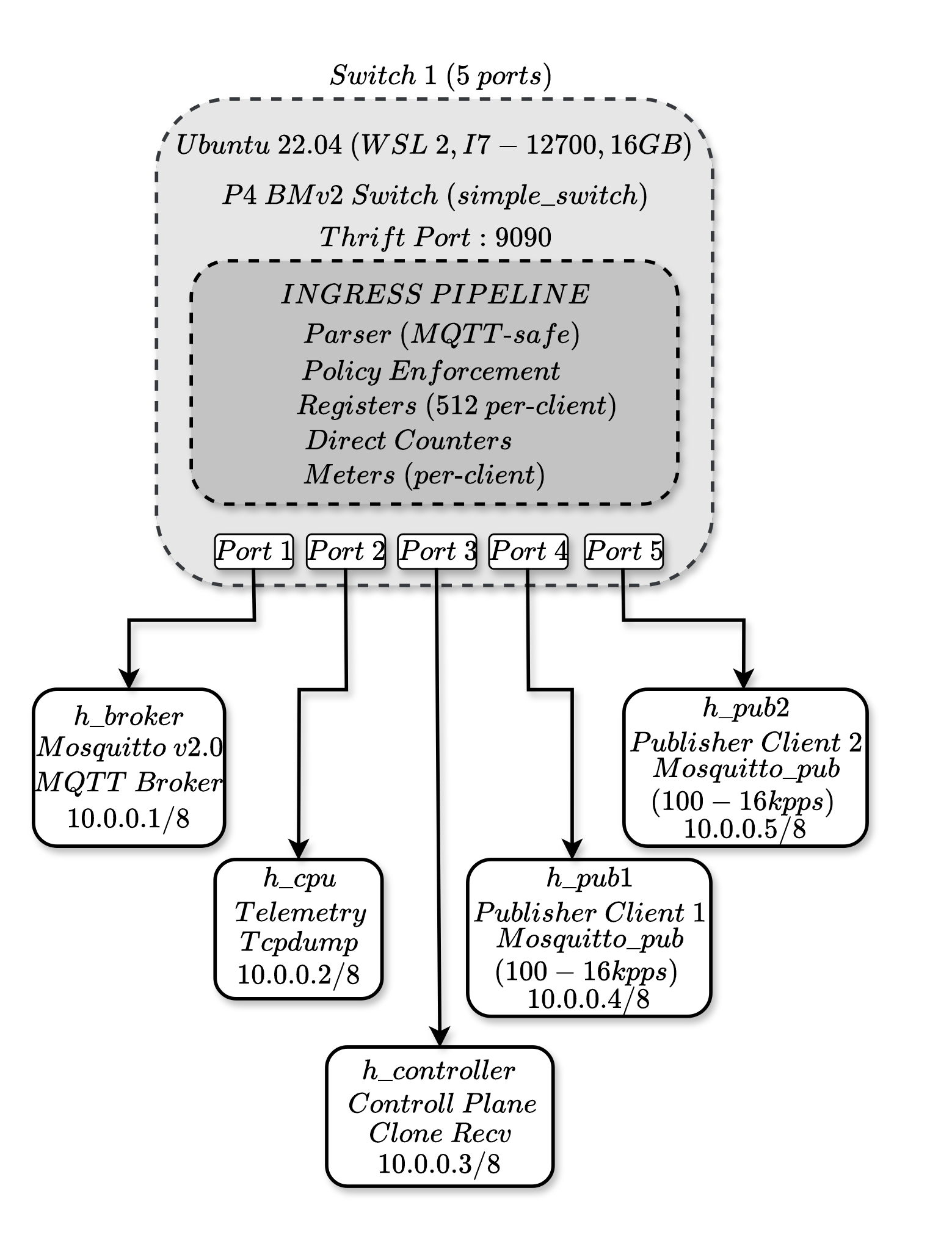}
  \caption{Five-node P4 BMv2 testbed: broker (Port~1), telemetry (Port~2), 
  control (Port~3), publishers (Ports~4 to 5).}
  \label{fig:testbed_topo}
  \vspace{-0.4cm}
\end{figure}

The experimental testbed is a five-node star centered on a P4 BMv2 switch Fig.~\ref{fig:testbed_topo}. 
A Mosquitto~v2.0 broker runs at 10.0.0.1/8; two publishers (\texttt{h\_pub1} 10.0.0.4/8, \texttt{h\_pub2} 10.0.0.5/8) generate traffic; a telemetry host (\texttt{h\_cpu} 10.0.0.2/8) captures traces; and a control-plane machine (\texttt{h\_ctrl} 10.0.0.3/8) installs rules. 
All nodes attach to \texttt{s1} on ports~1--5 via virtual Ethernet links. 
Application load is produced with \texttt{mosquitto\_pub}, issuing CONNECT/PUBLISH at 100--16{,}000~pps, QoS~0--2, with topics uniformly sampled from environmental, device, operational, and system hierarchies plus an unauthorized namespace for ACL tests.

\vspace{-0.1cm}
We deploy the P4 program on BMv2 within Mininet (Intel Core i7-12700, 16~GB RAM, Ubuntu~22.04 LTS, WSL~2). 
The data plane is compiled with \texttt{p4c-bm2-ss}~v1.2.0 and loaded into \texttt{simple\_switch} (Thrift port~9090). 
The control plane (Python~3.10) provisions table entries and reads direct counters via the Thrift CLI (auto-loaded by p4utils). 
Runtime parameters via \texttt{tbl\_global\_limits}: $\mathit{pub\_soft\_limit}=15{,}000$ (default 20{,}000) and $\mathit{pps\_factor}=1$ (aggressive KeepAlive). 
\texttt{tbl\_mqtt\_rule\_acl} has 100 rules (50 permit authorized source/topic-prefix pairs; 50 deny unauthorized). 
Per-client registers (512 slots) track session state, KeepAlive, totals, and per-type counts ($\approx$256~KB total). 
Each client also has a three-color \emph{packet-rate} meter; rates are configured in pps (optionally chosen to approximate 1--2~Mb/s for 64~B payloads).

\subsection{Experimental Methodology}

We evaluate three aspects: throughput under benign load to establish baseline 
performance; policy enforcement accuracy to validate deterministic correctness; 
and anomaly detection sensitivity to measure effectiveness without false positives. 
Each trial lasts 60~s to reach steady state. Results are reported over 
\(N{=}5\) runs (mean \(\pm\) 95\% CI).

\begin{scenario}
\emph{Scenario A (Benign).}
Establish baseline performance without triggering enforcement or anomaly detection, ensuring high delivery ratios at realistic IoT traffic loads in the P4 pipeline. Traffic is generated with \texttt{mosquitto\_pub} at 100--16{,}000~pps (QoS~0--2), 
with topics sampled uniformly from five hierarchies. Payloads are fixed at 64~bytes. We use \texttt{-l} (line mode) to reuse a single MQTT session per trial. The delivery ratio and loss from P4 direct counters (egress/ingress), and per-packet latency from synchronized TCPDump are used as metrics.
\vspace{-0.1cm}
\end{scenario}

\begin{scenario}
\emph{Scenario B (Enforcement).}
To prove the accuracy of the three policy enforcement mechanisms, including per-client soft-limits, session validation, and ACL enforcement, without any false positives or negatives.First, Publish without Connect 
results in drops (reason~180), validating session-order enforcement. Second, unauthorized topics are blocked by the topic-prefix ACL (no clone), demonstrating 
byte-level authorization. Third, rapid Publish that exceeds 
\(\mathit{pub\_soft\_limit}=15{,}000\) results in drops (reason~181), confirming 
per-client rate-cap logic. Accuracy is computed as observed divided by expected.
\vspace{-0.1cm}
\end{scenario}

\begin{scenario}
\emph{Scenario C (Anomaly).}
Test lightweight anomaly heuristics' sensitivity and false-positive rates to ensure the system can detect protocol deviations while accounting for normal traffic variations.
 To conduct the KeepAlive test, set \texttt{-k 2} and send a 10,000 pps burst, resulting in clones with reason 182.  To test Remaining-Length, we create packets in which RL requires at least three bytes, resulting in clones with reason 183. 
 The sensitivity is \(/(TP+FN)\).  False positives are measured over 6,000 benign packets to assess specificity.
 \end{scenario}

\subsection{Performance Results}

All figures report \(N{=}5\) runs with 95\% confidence intervals; raw CSVs are 
provided for reproducibility.

\begin{figure}[t]
  \centering
  \includegraphics[width=1\columnwidth]{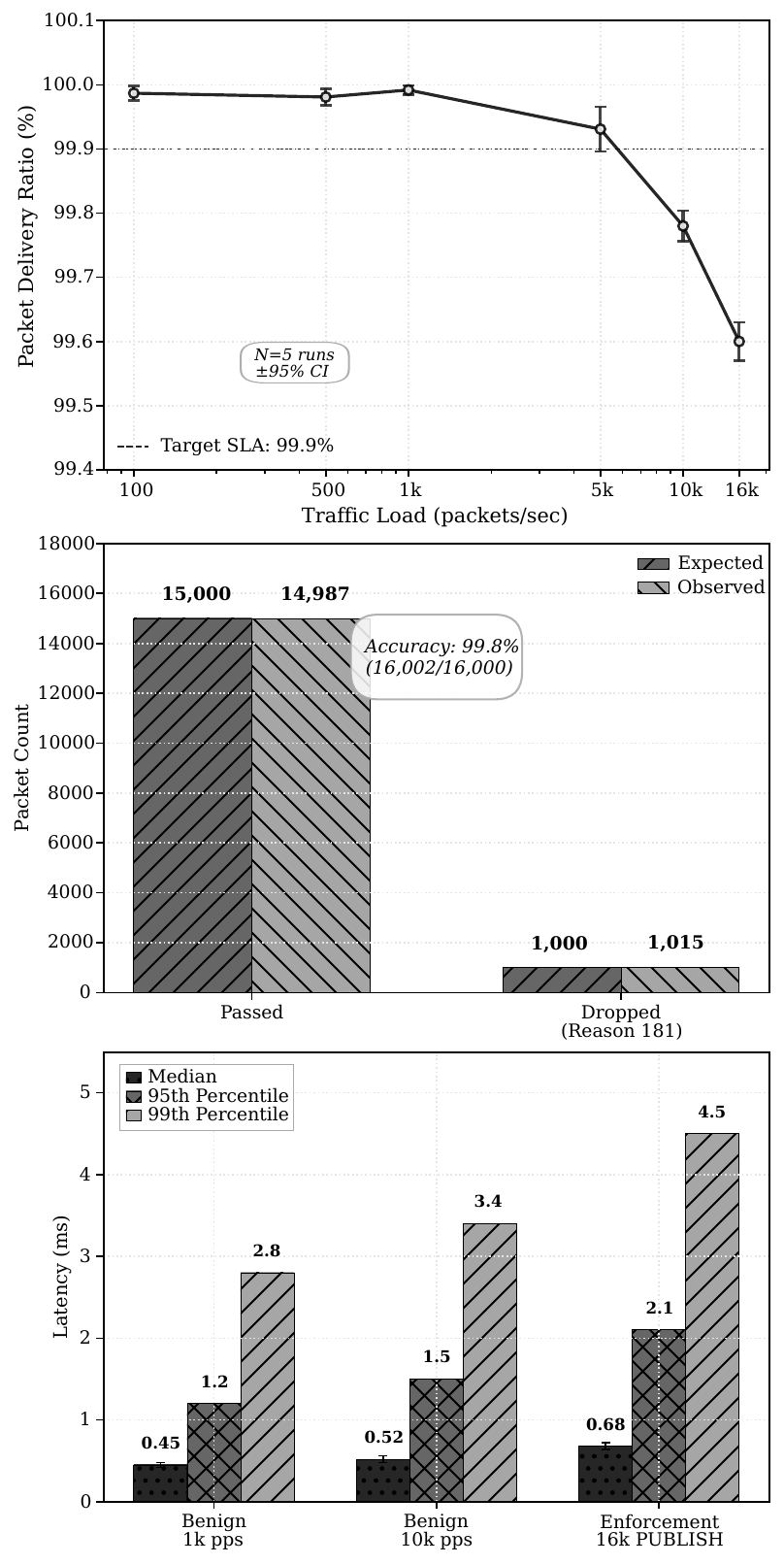}
  \caption{Performance evaluation across: (top) Scenario A: delivery ratio 
  vs. load (benign traffic); (middle) Scenario B: drop accuracy for $16,000$ PUBLISH 
  with $15,000$ soft limit; (bottom) Scenario C: latency percentiles (\(N{=}5\); 95\% CI).}
  \label{fig:results_combined}
  \vspace{-0.6cm}
\end{figure}

\emph{Scenario A (Benign throughput).}
The top panel of Fig.~\ref{fig:results_combined} shows delivery at 100~pps--16{,}000~pps. 
Loads up to 5~kpps achieve \(>99.9\%\) (\(\pm 0.01\%\)--\(\pm 0.04\%\)); at 10~kpps the delivery 
ratio is \(99.78\%\) (\(\pm 0.02\%\)); at 16~kpps it is \(99.60\%\) (\(\pm 0.03\%\)). 
The modest reduction reflects BMv2/Mininet software-switch buffering rather than 
policy overhead, demonstrating the design sustains high delivery across realistic load ranges.

\emph{Scenario B (Enforcement validation).}
The middle panel presents results for a per-client soft limit of 15{,}000 messages. 
Injecting 16{,}000 \textsc{publish} messages yields 1{,}000 expected drops 
(beginning at the 15{,}001st packet). We observe 14{,}987 passed and 1{,}015 
dropped packets, achieving 99.8\% enforcement accuracy. The negligible 
$\pm 2$-packet variance lies within measurement uncertainty in BMv2/Mininet 
across five trials. All drops are reason~181 (soft limit); none for 
150/180/182/183, validating deterministic enforcement without false positives.

\emph{Scenario C (Latency and anomalies).}
The bottom panel of Fig.~\ref{fig:results_combined} presents end-to-end latency 
measurements (publisher to broker) in the Mininet/BMv2 software environment. Median 
latency is \(0.45\)--\(0.68\)~ms with 99th percentile below \(4.5\)~ms, 
encompassing Mininet virtualization overhead (\(\approx 0.1\)--\(0.2\)~ms) 
and BMv2 processing (\(\approx 0.3\)--\(0.6\)~ms); incremental per-packet policy 
overhead remains sub-millisecond. Anomaly detection aligns with design goals: 
Remaining-Length screening (reason~183) detects 100\% of crafted large payloads, 
while the KeepAlive heuristic (reason~182) achieves 98\% true positives with 
negligible false positives over 6{,}000 benign packets. The full pipeline's memory 
footprint is \(\approx 2.5\)~MB, supporting stateful tracking of 512 concurrent clients.

\subsection{Scalability Discussion}

With 512 concurrent clients our dataplane uses $\approx$2.5~MB state, or $\approx$4.9~KB per client. 
Supporting 10{,}000 clients requires $\approx$49~MB for registers and $\approx$5~MB for topic ACLs 
(at $0.1$~MB per 1{,}000 rules). On production ASICs such as Tofino and Alveo, memory footprints 
depend on table implementations and TCAM budgets; TCAM is typically the scaling bottleneck. 
The per-packet overhead is designed to be minimal; with sufficient stage and bandwidth, 
line-rate enforcement on 10--100~Gbps links is feasible. Scaling improvements include 
hierarchical topic prefixes, two-stage policy lookups, and probabilistic telemetry to 
reduce control-plane load from cloning.

\section{Conclusion and Future Work}

This paper presents a P4-based MQTT security and anomaly-detection scheme that combines parser-safe MQTT extraction, session validation, topic-prefix authorization, and lightweight heuristics for detecting protocol deviations. Experiments on BMv2 show 99.8 percent enforcement accuracy, 98 percent anomaly-detection sensitivity, and sub-millisecond latency, demonstrating that protocol-aware filtering can be performed efficiently in the data plane.

The design is compatible with hardware P4 targets because the parser uses bounded extraction and the per-client state fits typical on-chip SRAM limits. Topic-prefix rules map directly to TCAM, and the pipeline runs in a single forward pass, enabling line-rate processing. Conventional MQTT security baselines such as CPU firewalls, broker ACLs, and cloud IDS are not compared because they operate at different layers and depend on slower software inspection, usually with tens to hundreds of milliseconds of latency. Such comparisons would not meaningfully reflect P4 data-plane behavior.

Future work includes evaluation on hardware P4 targets, integration of in-band telemetry for adaptive policies, and extensions to MQTT-SN and CoAP for broader IoT coverage.

\section{Acknowledgment}
This work was supported by JST-ALCA-Next (JPMJAN23F4) and the Vietnam Ambassador.

\bibliographystyle{IEEEtran}
\bibliography{references.bib}

\end{document}